\documentclass[manuscript]{aastex}
\usepackage{amsmath}
\usepackage{wasysym}

\begin{document}

\title{Cosmic Microwave Background constraints of decaying dark matter particle properties}

\author{S. Yeung, M. H. Chan, and M. -C. Chu\altaffilmark{1}}
\affil{Department of Physics and Institute of Theoretical Physics,
The Chinese University of Hong Kong,
Shatin, New Territories, Hong Kong, China}

\altaffiltext{1}{Email address: mcchu@phy.cuhk.edu.hk}

\begin{abstract}
If a component of cosmological dark matter is made up of massive particles - such as sterile neutrinos - that decay with cosmological lifetime to emit photons, the reionization history of the universe would be affected, and cosmic microwave background anisotropies can be used to constrain such a decaying particle model of dark matter. The optical depth depends rather sensitively on the decaying dark matter particle mass $m_\mathrm{dm}$, lifetime $\tau_\mathrm{dm}$, and the mass fraction of cold dark matter $f$ that they account for in this model. Assuming that there are no other sources of reionization and using the WMAP 7-year data, we find that 250~eV $\apprle$ $m_\mathrm{dm}$ $\apprle$ 1 MeV, whereas $2.23\times 10^{3}$~yr $\apprle$ $\tau_\mathrm{dm}/f$ $\apprle$ $1.23\times 10^{18}$~yr. The best fit values for $m_\mathrm{dm}$ and $\tau_\mathrm{dm}/f$ are 17.3 keV and $2.03\times 10^{16}~\mathrm{yr}$ respectively.
\end{abstract}


\section{Introduction}
There have been tremendous progress in cosmology in the past decade.   The availability of high quality observational data such as those from WMAP \citep{kom10} has led to tight constraints on cosmological parameters and models.  There is now a standard model of cosmology, in which only a small portion of the total mass-energy in the universe is ordinary matter, the rest being dark components which we have little understanding of.  In some of the proposed dark matter models, such as sterile neutrino \citep{dol94}, the dark matter particles may decay and emit photons \citep{bor08, cen01}, which are redshifted with the expansion of the universe and may eventually ionize hydrogen and helium at later times.  Therefore, decaying dark matter particles may contribute to reionization and imprint their signatures on the cosmic microwave background anisotropies (CMBA). In this paper, we constrain the mass and life time of decaying dark matter particles by using the WMAP data of CMBA.

There are strong evidences for reionization at late universe \citep{bec01}, many sources of which have been proposed, such as star formation \citep{gne97}, UV radiation from black holes \citep{sas96}, and supernova-driven winds \citep{teg93}. \citet{bie06} point out that the X-ray photons produced in the decays of sterile neutrinos can boost the production of molecular hydrogen, and as a result the rates of cooling of gas and early star formation are increased, leading to reionization at redshift consistent with the WMAP results. \citet{boy06} use extragalactic diffuse X-ray background to constrain the decay rate of sterile neutrinos as a warm dark matter candidate. \citet{sel06} use the Ly-alpha forest power spectrum measured by the Sloan Digital Sky Survey and high-resolution spectroscopy observations in combination with WMAP data and galaxy clustering to constrain sterile neutrino masses.  The lower limits obtained are 13.1 keV at 95\% C.L. and 9.0 keV at 99.9\% C.L.. In \citet{zha07}, decaying dark matter is also considered to be an energy source of reionization. However several approximations are made in that paper; in particular, the fraction of the decay energy deposited in baryonic gas is simply characterized by a phenomenological parameter. In this paper, we do not make approximations about the amount of energy absorbed by the baryons. The ionization and heating rates are calculated using the appropriate cross sections. Furthermore, we vary both the decaying dark matter particle parameters and cosmological parameters to fit the CMBA data, while in \citet{zha07} only the decaying dark matter particle parameters and the scalar amplitude are varied to fit the CMBA spectrum. There is also another earlier work \citep{map05} studying the effect on reionization by the decaying process of sterile neutrinos, where a relation between sterile neutrino mass and lifetime is used, which is based on the assumption that sterile neutrinos are the dominant component of dark matter. However in this paper we do not assume any relation between the decaying sterile neutrino mass and lifetime \footnote{The lifetime refers to the radiative channel only. Since the sterile neutrinos can also decay into 3 active neutrinos, the total lifetime $\approx \tau_\mathrm{dm}/128$ \citep{bar95}.}; we treat them as two independent parameters instead, and we introduce another free parameter, the mass fraction of dark matter that are decaying, $f$.

The optical depth depends rather sensitively on the mass $m_\mathrm{dm}$ and life time $\tau_\mathrm{dm}$ of dark matter particles, as well as $f$. However, to good approximation, the effects of $f$ and $\tau_\mathrm{dm}$ are degenerate and only their ratio is an independent parameter. We then constrain these parameters, $m_\mathrm{dm}$ and $\tau_\mathrm{dm}/f$ by the WMAP 7-year data. Assuming that such a decaying process is the only source of reionization \footnote{There is a well-known discrepancy between reionization redshifts deduced from CMBA and quasar absorption line observations. However, the constraints based on quasar absorption line observations are highly model-dependent \citep{mcg11}.  In particular, the steep rise in the Gunn-Peterson effective optical depth at z $\apprge$ 6 is highly controversial, as it is very sensitive to the assumed density field and continuum fitting \citep{bec07}. Recently, direct and  model-independent limits on the fraction of neutral Hydrogen at z $\approx$ 5-6 were obtained using the simple statistic of the covering fraction of dark pixels, and they can be consistent with the ionization history derived from CMBA observations \citep{mcg11}. On the other hand, a recent work shows that model independent joint CMBA-quasar absorption line constraints still permit a broad range of reionization history for $z > 6$ \citep{mit12}.}, we find that $m_\mathrm{dm}$ is less than about 1 MeV, and $\tau_\mathrm{dm}/f$ is less than about $10^{18}~\mathrm{yr}$, with the best-fit values being 17.3 keV and $2.03\times 10^{16}~\mathrm{yr}$ respectively.

In Section 2, we present the calculation of the ionization fraction and optical depth in this decaying dark matter model. The Markov Chain Monte Carlo fitting to WMAP data results and discussion are presented in Section 3, and Section 4 is a summary and conclusion.

\section{The Model}

The evolution of the ionization fractions $x_\mathrm{H}(z)$ and $x_\mathrm{He}(z)$ (for hydrogen and helium respectively) and the matter temperature $T_m(z)$ satisfy the coupled ordinary differential equations
\begin{align}
\frac{dx_\mathrm{H}(z)}{dz}=\frac{-1}{(1+z)n_\mathrm{H}(z)H(z)}[R_{i\mathrm{H}}(z)+R_{sx\mathrm{H}}(z)]\label{ode1}
\\
\frac{dx_\mathrm{He}(z)}{dz}=\frac{-1}{(1+z)n_\mathrm{He}(z)H(z)}[R_{i\mathrm{He}}(z)+R_{sx\mathrm{He}}(z)]\label{ode2}
\\
\frac{dT_m(z)}{dz}=\frac{-2}{3k_\mathrm{B}(1+z)H(z)}\frac{R_{h\mathrm{H}}(z)+R_{h\mathrm{He}}(z)+R_{sT}(z)}{[1+x_\mathrm{H}(z)+x_\mathrm{He}(z)]n_\mathrm{H}(z)},\label{ode3}
\end{align}
where $R_{sx}(z)$ and $R_{sT}(z)$ are the standard net recombination and net heating rates respectively, and $R_{iS}(z)$ and $R_{hS}(z)$ are the additional ionization and heating rates respectively due to the decaying dark matter particles. The additional terms include the contribution from both hydrogen ($S=\mathrm{H}$) and helium ($S=\mathrm{He}$) atoms.

The hydrogen ionization rate $R_{i\mathrm{H}}(z)$ due to the decaying dark matter particles is:
\begin{equation}\label{ionequi}
R_{i\mathrm{H}}(z) = n_\mathrm{H}(z) \left[1-x_\mathrm{H}\left(z\right)\right]
\int^\infty_{E_\mathrm{th,H}} \frac{4\pi J(E)}{E}
\sigma_\mathrm{H}\left(E\right) dE,
\end{equation}
where $x_\mathrm{H}\left(z\right)$ is the ionization fraction at redshift $z$, 
$n_\mathrm{H}(z)$ is the total hydrogen number density including neutral and ionized hydrogen at redshift $z$, 
$J\left(E\right)$ is the photon energy flux per unit energy per unit solid angle at energy $E$, and $\sigma_\mathrm{H}\left(E\right)$ is the photoionization cross section of hydrogen at energy $E$. The hydrogen heating rate $R_{h\mathrm{H}}(z)$ due to the decaying dark matter particles is:
\begin{equation}\label{heat_eqn1}
R_{h\mathrm{H}}(z) = n_\mathrm{H}[1-x_\mathrm{H}(z)]\int^\infty_{E_\mathrm{th,H}} \frac{4\pi J(E)(E-E_\mathrm{th,H})}{E}
\sigma_\mathrm{H}\left(E\right) dE.
\end{equation}
Similarly, the helium ionization rate $R_{i\mathrm{He}}(z)$ due to the decaying dark matter particles is:
\begin{equation}\label{ionequiHe}
R_{i\mathrm{He}}(z) = n_\mathrm{He}(z) \left[1-x_\mathrm{He}\left(z\right)\right] 
\int^\infty_{E_\mathrm{th,He}} \frac{4\pi J(E)}{E}
\sigma_\mathrm{He}\left(E\right) dE,
\end{equation}
and the helium heating rate $R_{h\mathrm{He}}(z)$ due to the decaying dark matter particles is:
\begin{equation}\label{heat_eqn1He}
R_{h\mathrm{He}}(z) = n_\mathrm{He}[1-x_\mathrm{He}(z)]\int^\infty_{E_\mathrm{th,He}} \frac{4\pi J(E)(E-E_\mathrm{th,He})}{E}
\sigma_\mathrm{He}\left(E\right) dE.
\end{equation}
The photon flux $J(E)$ is
\begin{equation}
J(E)=\frac{1}{4\pi}\left[n_\mathrm{dm0}(1+z)^3\right]\frac{c~e^{-t(z_\mathrm{em})/\tau_\mathrm{dm}}e^{-\tau_\mathrm{abs}(z,z_\mathrm{em})}}{H(z_\mathrm{em})\tau_\mathrm{dm}},
\label{JE} 
\end{equation}
where $z_\mathrm{em}$ is the red shift when the dark matter particle decayed,  $E=E_0\frac{1+z}{1+z_\mathrm{em}}$, $E_0$ being the energy of the emitted photon, $t(z)$ is the time elapsed since the big bang at redshift $z$, $n_\mathrm{dm0}$ is the present number density of the decaying dark matter particles if they did not decay so that $n_\mathrm{dm0}(1+z)^3$ is the number density at redshift $z$, and $\tau_\mathrm{abs}$ is the optical depth caused by absorption of photons by ionization defined by
\begin{equation}\begin{split}
\label{eq:tau_abs}
\tau_\mathrm{abs}(z,z_\mathrm{em})\equiv\int^{z_\mathrm{em}}_z \frac{1}{(1+z^\prime)H(z^\prime)}\left[1-x_\mathrm{H}\left(z^\prime\right)\right]n_\mathrm{H}(z^\prime)\sigma_\mathrm{H}(E=E_0\frac{1+z^\prime}{1+z_\mathrm{em}})cdz^\prime\\
+\int^{z_\mathrm{em}}_z \frac{1}{(1+z^\prime)H(z^\prime)}\left[1-x_\mathrm{He}\left(z^\prime\right)\right]n_\mathrm{He}(z^\prime)\sigma_\mathrm{He}(E=E_0\frac{1+z^\prime}{1+z_\mathrm{em}})cdz^\prime.
\end{split}\end{equation}

We consider photon energy $E$ between the hydrogen threshold energy $E_\mathrm{th,H}$(=13.6 eV) and $E_0$ at redshift $z$. The lower limit is $E_\mathrm{th,H}$ since photons with $E$ below $E_\mathrm{th,H}$ do not have enough energy to ionize hydrogen. The upper limit is $E_0$ since the redshift due to the expansion of the universe would cause photons to have energy $E$ smaller than $E_0$. Hence the integral in (\ref{ionequi}) can be written as:
\begin{align}
&\int^\infty_{E_\mathrm{th,H}} \frac{4\pi J(E)}{E}
\sigma_\mathrm{H}\left(E\right) dE\nonumber
\\
&=\frac{c~n_\mathrm{dm0}(1+z)^3}{\tau_\mathrm{dm}}\int^{E_0}_{E_\mathrm{th,H}} \frac{e^{-t(z_\mathrm{em})/\tau_\mathrm{dm}}e^{-\tau_\mathrm{abs}(z,z_\mathrm{em})}}{E~H(z_\mathrm{em})}
\sigma_\mathrm{H}\left(E\right) dE\nonumber
\\
&=\frac{c~n_\mathrm{dm0}(1+z)^3}{\tau_\mathrm{dm}}\int^{E_0/(1+z)}_{E_\mathrm{th,H}/(1+z)} \frac{e^{-t(z_\mathrm{em}=E_0/E_\mathrm{obs}-1)/\tau_\mathrm{dm}}e^{-\tau_\mathrm{abs}(z,z_\mathrm{em}=E_0/E_\mathrm{obs}-1)}}{E_\mathrm{obs}H(z_\mathrm{em}=E_0/E_\mathrm{obs}-1)}
\sigma_\mathrm{H}\left(E_\mathrm{obs}(1+z)\right) dE_\mathrm{obs}\label{eq:ion_int},
\end{align}
where we have made a change of integration variable in the third line from $E$ to $E_\mathrm{obs}$, the present observed energy of a photon produced in the past by the decaying process. The integral for the helium contribution can be rewritten similarly. The photoionization cross section $\sigma(E)$ is approximated by \citet{astrophy-book}:
\begin{equation}
\sigma(E)=\sigma_\mathrm{th} \left\{ \beta\left[ \frac{E}{E_\mathrm{th}}\right] ^{-s}+(1-\beta)\left[ \frac{E}{E_\mathrm{th}}\right] ^{-(s+1)} \right\},
\end{equation}
where $\sigma_\mathrm{th}=6.30\times 10^{-18} \mathrm{cm}^2$, $\beta=1.34$ and $s=2.99$ for hydrogen, and $\sigma_\mathrm{th}=7.42\times 10^{-18} \mathrm{cm}^2$, $\beta=1.66$ and $s=2.05$ for helium. To account for the energetic secondary electrons that could ionize and heat up hydrogen and helium atoms, we multiply the cross sections in the ionization rates by an additional factor $\{1+\phi[x(z)]E(z)/E_{\rm th}\}$ \citep{map05}. For the cross sections in the ionization rates, $\phi(x)=C(1-x^a)^b$ and $C=0.3908$, $a=0.4092$ and $b=1.7592$ for hydrogen, and $C=0.0554$, $a=0.4614$ and $b=1.6660$ for helium \citep{shu85}. For the cross sections in the heating rates,  $\phi(x)=C[1-(1-x^a)^b]$ and $C=0.9971$, $a=0.2663$ and $b=1.3163$ for hydrogen and helium. In this model we take the Hubble parameter $H(z)$ to be that given in the $\Lambda$CDM model, $H_0\sqrt{\Omega_\Lambda+\Omega_m(1+z)^3}$ where $H_0 = H(z = 0)$ is the present Hubble parameter, $\Omega_m$ is the present matter density, and $\Omega_\Lambda$ is the dark energy density.

There are three free parameters $m_\mathrm{dm}$, $\tau_\mathrm{dm}$, $f$ in this model, where 
$m_\mathrm{dm}$ and $\tau_\mathrm{dm}$ are the mass and life time of the decaying dark matter particles, and
$f$ is the present mass fraction of the total dark matter accounted for by these hypothetical decaying particles. $m_\mathrm{dm}$ and $f$ are directly related to $E_0$ and $n_\mathrm{dm0}$ in the above equations:
\begin{eqnarray}
&E_0=m_\mathrm{dm}/2&,
\\
&n_\mathrm{dm0}=f\Omega_\mathrm{dm} \frac{3H_0^2}{8\pi G m_\mathrm{dm}c^2}e^{t_0/\tau_\mathrm{dm}}&,\label{params_ndm0}
\end{eqnarray}
where $\Omega_\mathrm{dm}$ is the present dark matter density, $G$ is the gravitational constant, and $t_0$ is the age of the universe. If we assume $\tau_\mathrm{dm}$ is much larger than $t_0$, the exponential factors in Eqs.~(\ref{eq:ion_int}) and (\ref{params_ndm0}) can be ignored. The ionization and heating rates due to decaying dark matter particles then depend only on the ratio between $f$ and $\tau_\mathrm{dm}$.

By solving the two ordinary differential equations (\ref{ode1}) and (\ref{ode2}) we get the ionization fraction $x$ and matter temperature $T_m$ as functions of $z$. The subprogram RECFAST \citep{sea00} used in CAMB \citep{lew00} is modified to include the additional terms due to the decaying dark matter particles. We have changed CAMB so that the reionization history $x(z)$ is calculated by our model, rather than the \textit{ad hoc} $\tanh$ function built in. This introduces three new parameters: $f$, $\tau_\mathrm{dm}$ and $m_\mathrm{dm}$, which are varied together with other cosmological parameters. We have not changed the Boltzmann equations since the effects introduced by the decaying cold dark matter are very small and are of higher order. For example, the photons from the dark matter decay affect other photons by first scattering with electrons which then scatter with other photons. Also, very few dark matter particles have decayed by the time of recombination. The fraction of electrons affected by the decay photons can be approximated by the ratio between the number densities of the decay photons and the CMB photons, which is at most of the order $10^{-7}$ within the interested ranges of $f$ and $\tau_{\rm dm}$. Therefore only very few electrons and hence CMB photons are affected by the decaying dark matter particles directly. COSMOMC \citep{lew02} and the 7-year WMAP data are used to constrain the parameters in this model.

\section{MCMC Results and Discussion}

\begin{figure}
\plottwo{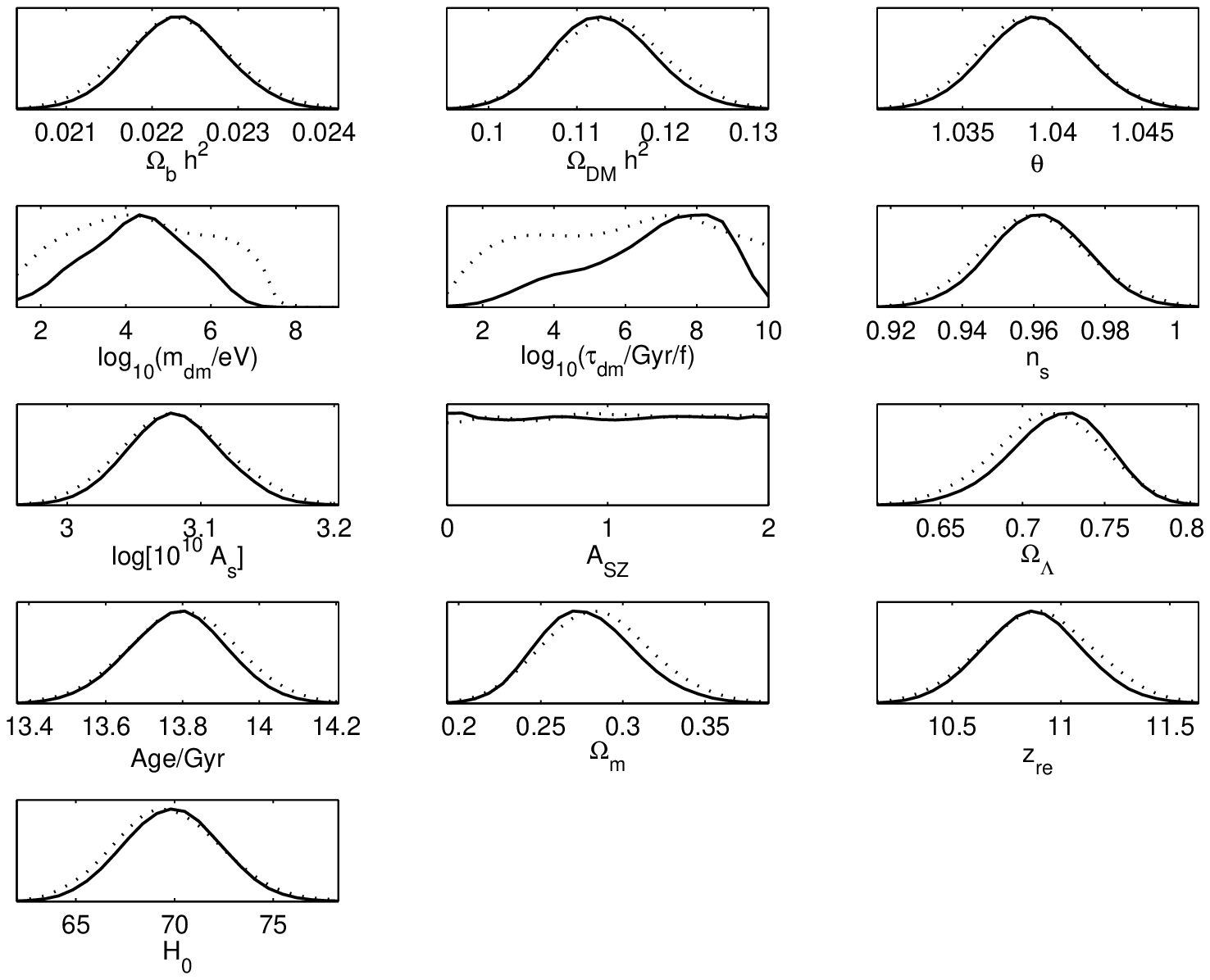}{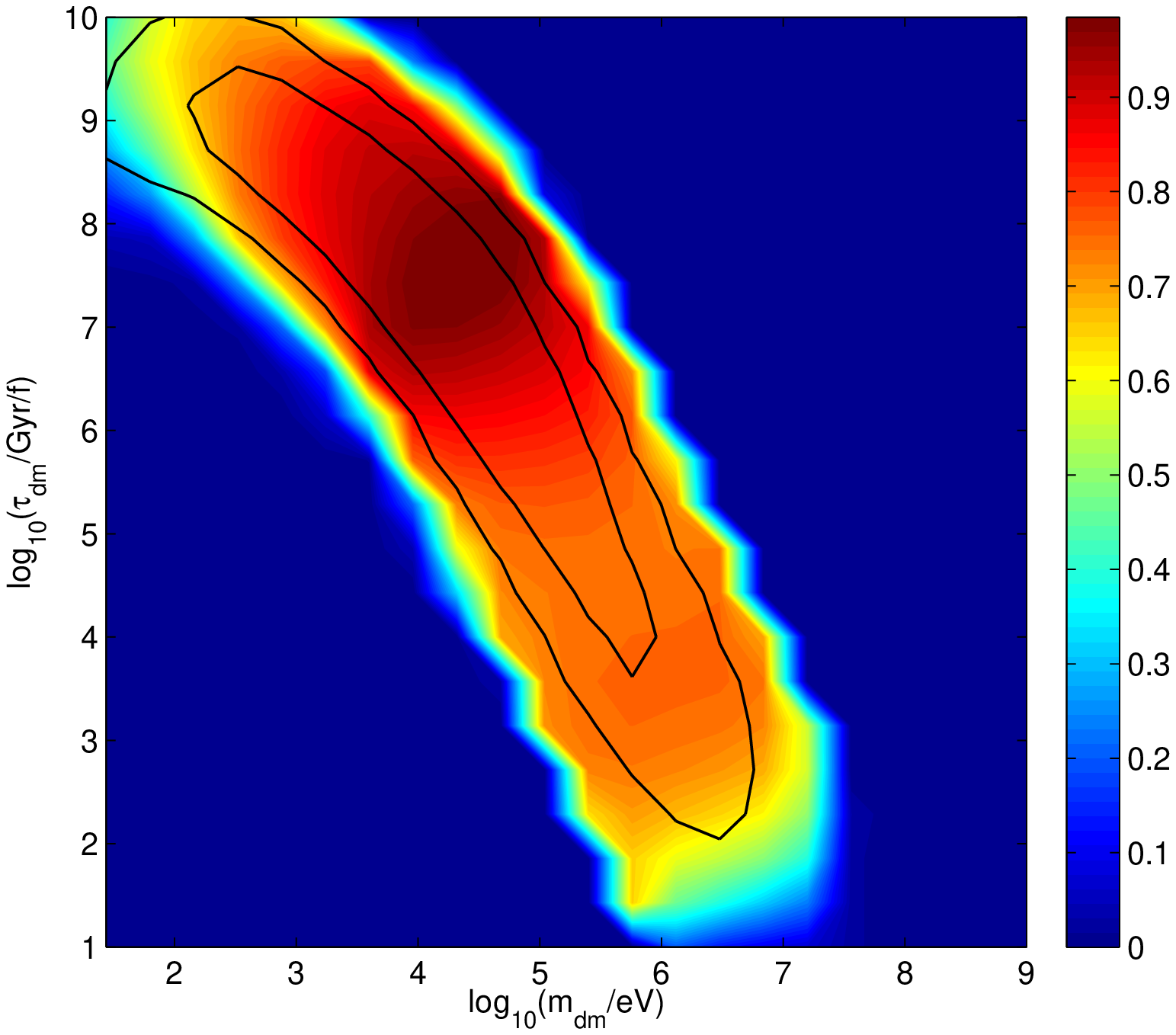}
\caption{Left: Probability distribution of the parameters in the decaying dark matter model. The solid line is the marginalized likelihood and the dotted line is the mean likelihood. Right: Joint constraints of the three dark matter parameters in the decaying dark matter model. The contours refer to the
marginalized likelihoods while the colours refer to the mean likelihood. The inner contour corresponds to 68\% confidence limit and the outer contour corresponds to 95\% confidence limit. Relative values of the likelihood are used, with the maximum likelihood set to be 1, and the color bar shows the relation between the colors and the relative likelihood. The 7-year WMAP results are used to constrain the parameters. Refer to Table \ref{xbestfit} for explanation of the parameters. $\theta$ is 100 times the ratio of the sound horizon to the angular diameter distance at about the time of recombination, and $A_\mathrm{SZ}$ is the floating amplitude for the SZ spectrum as in WMAP.}
\label{fig:cosmomc_results}
\end{figure}

By varying $m_\mathrm{dm}$, $\tau_\mathrm{dm}$, $f$ and other standard parameters in COSMOMC and comparing the resulting CMBA spectrum with WMAP 7-year data, we obtain the constraints on these parameters. The lower limits of $m_\mathrm{dm}$, $\tau_\mathrm{dm}$ and $\tau_\mathrm{dm}/f$ are set to be 13.6 eV, $10^{10}~\mathrm{yr}$ and $10^{10}~\mathrm{yr}$ respectively. The minimum value for $\tau_\mathrm{dm}$ is chosen to be 10~Gyr because it has to be large enough so that not too much dark matter have already decayed by now. The result is shown in Figure~\ref{fig:cosmomc_results}.

From the COSMOMC results the set of best fit parameters is shown in Table \ref{xbestfit}, and the marginalized limits are shown in Table \ref{xlimits}.

\begin{table}[htbp]
\centering
\caption{Best fit parameters in the decaying dark matter model. $A_s$ is defined at the pivot scale of 0.05 Mpc$^{-1}$.}
\label{xbestfit}
\begin{tabular}{|c|c|c|c|c|c|c|}
\hline
Parameter &Symbol &Value\\ \hline
Hubble parameter ($\mathrm{km}~\mathrm{s}^{-1} \mathrm{Mpc}^{-1}$) &$H_0$ &$69.6$\\
Physical baryon density &$\Omega_bh^2$ &$0.02238$\\
Physical dark matter density &$\Omega_ch^2$ &$0.1141$\\
Curvature fluctuation amplitude &$A_s$ &$2.17 \times 10^{-9}$\\
Scalar spectral index &$n_s$ &$0.963$\\
Decaying dark matter particle mass ($\mathrm{keV}$) &$m_\mathrm{dm}$ &$17.3$\\
Decaying dark matter particle life time over fraction (yr) &$\tau_\mathrm{dm}/f$ &$2.03\times 10^{16}$\\
\hline
\end{tabular}
\end{table}

\begin{table}[htbp]
\centering
\caption{List of the marginalized limits for different parameters in the decaying dark matter model, together with standard cosmological parameters.}
\label{xlimits}
\begin{tabular}{|c|c|c|c|c|c|c|}
\hline
Symbol &Prior &Limits (68\%) &Limits (95\%)\\ \hline
$H_0$ ($\mathrm{km}~\mathrm{s}^{-1} \mathrm{Mpc}^{-1}$) &\{40, 100\}&\{67.5, 72.3\} &\{65.3, 74.6\}\\
$\Omega_bh^2$ &\{0.005, 0.1\}&\{0.02176, 0.02284\} &\{0.02123, 0.02339\}\\
$\Omega_ch^2$ &\{0.01, 0.99\}&\{0.1076, 0.1185\} &\{0.1024, 0.1243\}\\
$\ln(10^{10}A_s)$ &\{2.7, 4\}&\{3.047, 3.115\} &\{3.013, 3.149\}\\
$n_s$ &\{0.5, 1.5\}&\{0.948, 0.975\} &\{0.935, 0.987\}\\
$\log_{10}(m_\mathrm{dm}/\mathrm{eV})$ &\{1.13, 9\}&\{3.79, 4.84\} &\{2.41, 6.05\}\\
$\log_{10}(\tau_\mathrm{dm}/f/\mathrm{Gyr})$ &\{1, 10\}&\{6.10, 7.90\} &\{3.36, 9.09\}\\
\hline
\end{tabular}
\end{table}

\begin{figure}
\plottwo{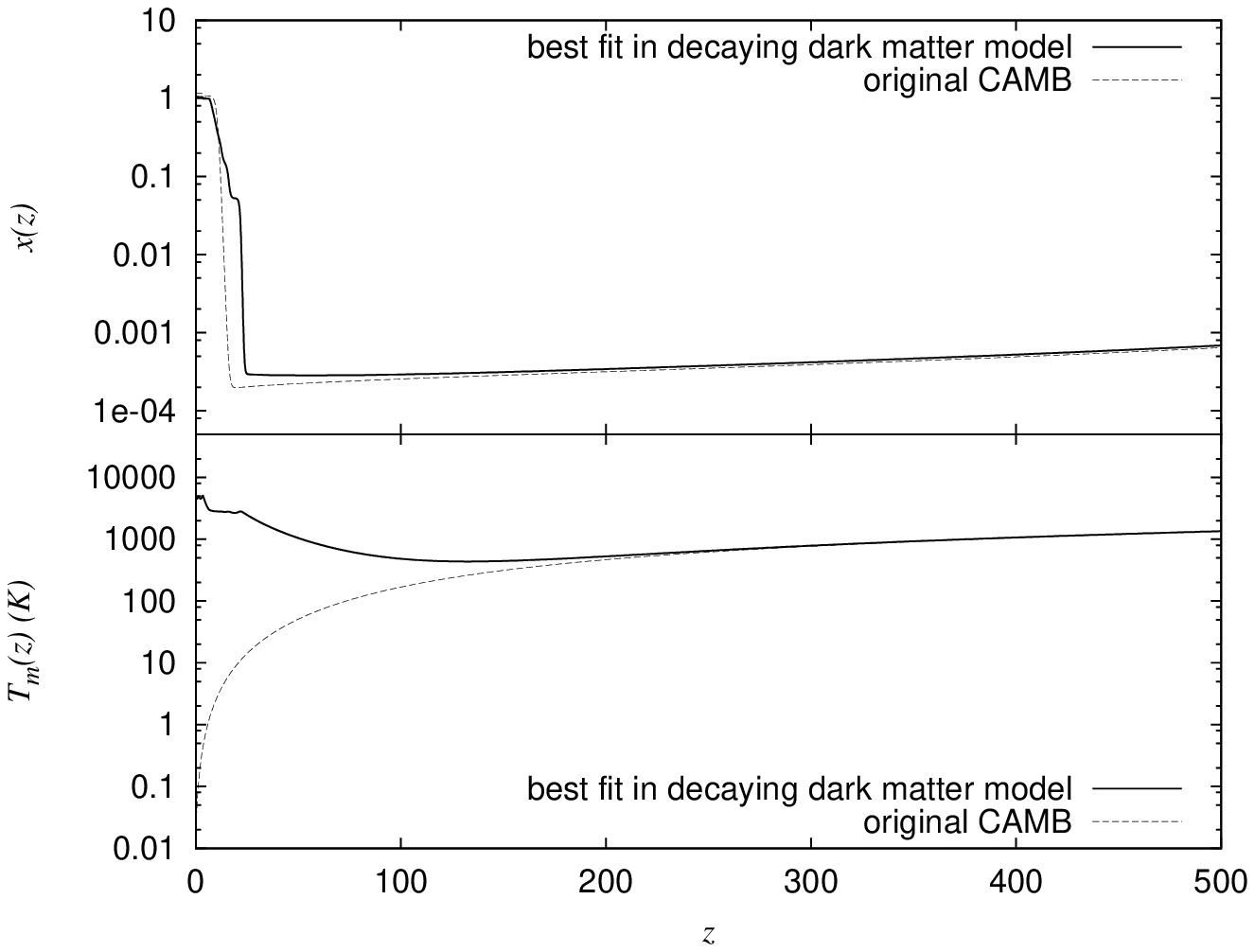}{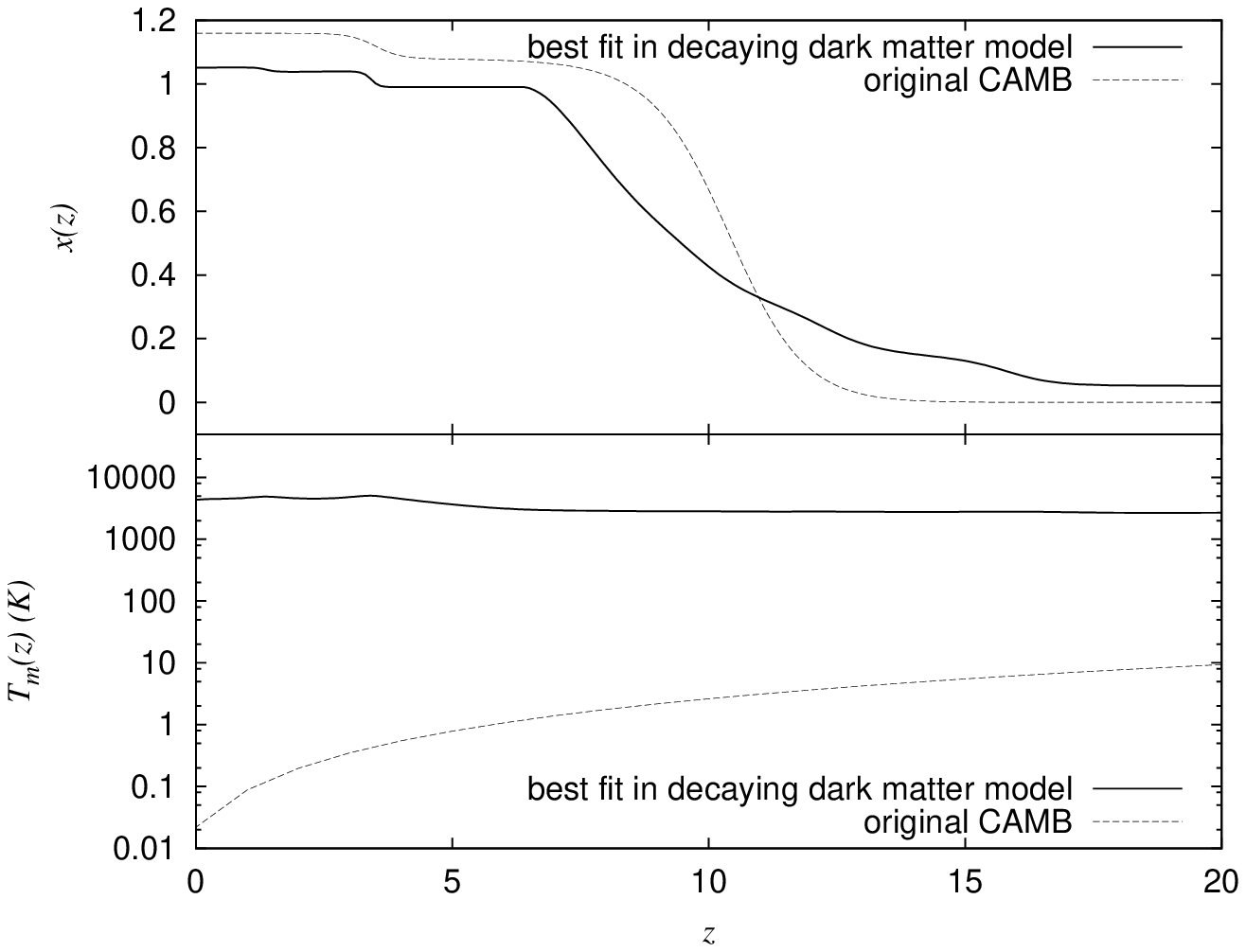}
\caption{Left: The upper figure plots the ionization fraction $x(z)$ versus $z$, and the lower figure plots the matter temperature $T_m(z)$ versus $z$, both using the decaying dark matter model with the best fit parameters in Table \ref{xbestfit}, and the original CAMB with the best fit WMAP parameters. Right: Same as left frame, but for the region around $z\approx$ 10.}
\label{fig:xe-fit}
\end{figure}

\begin{figure}
\centering
\includegraphics[width=1\linewidth]{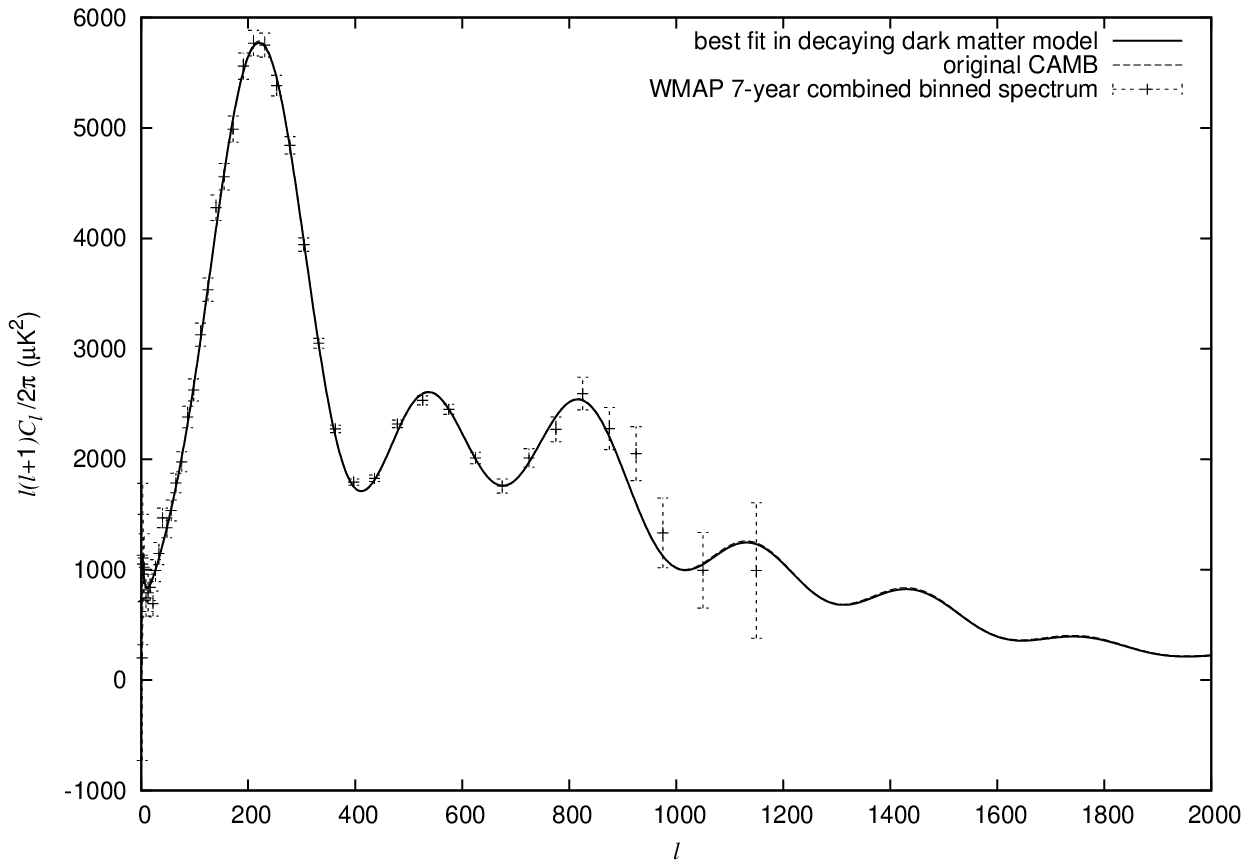}
\caption{CMBA temperature power spectra, calculated using the decaying dark matter model with the best fit parameters in Table \ref{xbestfit}, and the original CAMB with the best fit WMAP parameters. The values of the best fit $-\ln(\mathrm{likelihood})$ in the decaying dark matter model and standard $\Lambda$CDM model are 5527.44 and 5532.39 respectively.}
\label{fig:cl-fit}
\end{figure}

We can see that $m_\mathrm{dm}$ and $\tau_\mathrm{dm}/f$ are highly correlated, and are less than about 1 MeV and $10^{19}$~yr respectively. This mass range overlaps with that of sterile neutrinos in some models, for example \citet{boy06}. For comparison, the ionization fraction using the set of best fit parameters in the decaying dark matter model and assumed in the original CAMB are plotted together in Figure~\ref{fig:xe-fit}, and the corresponding CMBA spectra are plotted in Figure~\ref{fig:cl-fit}. From Figure~\ref{fig:xe-fit} we can see that the two different ionization fractions agree quite well for most values of $z$, but the difference is significant for $z\approx 50$. Nevertheless the resulting CMBA spectra are still nearly the same. We can also see that the ionization fraction in the decaying dark matter model actually agrees quite well with the ad hoc imposed $\tanh$ function in default CAMB. Recently, joint CMBA-quasar absoprtion line constraints on the reionization history using a model independent principal component decomposition method suggests that
reionization is 50\% complete between $9.0 < z < 11.8$, and 99\% complete between $5.8 < z < 10.4$ (95\% CL) \citep{mit12}. A similar study obtained a best-fit reionization history very close to our result presented in Figure~\ref{fig:xe-fit} (Figure~1 in \citet{pan11}). Another recent work based on the patchy kinetic Sunyaev-Zel'dovich effect concludes that reionization ended at $z > 5.8$ or 7.2 (95\% CL), depending on whether correlation with the cosmic infrared background is assumed or not \citep{zah11}. Our result shown in Figure~\ref{fig:xe-fit} is consistent with these recent constraints.

Our results are consistent with the constraint from diffuse X-ray background \citep{boy06}. The empirical bound in \citet{boy06} is $\log_{10}(\tau_\mathrm{dm}/\mathrm{Gyr}/f)$ $\apprge$ 8.5, which gives further reduction of the allowed region in the contour plot in Figure~\ref{fig:cosmomc_results}.

\section{Summary and Conclusion}

We have investigated the effects on CMBA by a component (mass fraction $f$) of dark matter particles with mass $m_\mathrm{dm}$ that decay with cosmological lifetime $\tau_\mathrm{dm}$. The photons emitted are redshifted and may ionize hydrogen and helium at later times, affecting the reionization history of the universe. If $\tau_\mathrm{dm}$ is much longer than the age of the universe, the optical depth depends only on the ratio of $\tau_\mathrm{dm}$ and $f$. We obtained constraints on these parameters by using the WMAP 7-year data and modified RECFAST, CAMB and COSMOMC codes and assuming that the only reionization source is the decaying dark matter. In the long lifetime limit, we find that 250~eV $\apprle$ $m_\mathrm{dm}$ $\apprle$ 1 MeV, $2.23\times 10^{3}$~yr $\apprle$ $\tau_\mathrm{dm}/f$ $\apprle$ $1.23\times 10^{18}$~yr, and the best fit values of $m_\mathrm{dm}$ and $\tau_\mathrm{dm}/f$ are 17.3 keV and $2.03\times 10^{16}~\mathrm{yr}$ respectively. Sterile neutrinos with mass 17.4~keV are possible within our marginal limits at 95\% CL, which may account for the 8.7~keV emission observed by the $Suzaku$ mission \citep{cha10, pro10}. The allowed range of $\tau_\mathrm{dm}/f$ is reduced further if the constraint from diffuse X-ray background is taken into account: $3.16\times 10^{17}$~yr $\apprle$ $\tau_\mathrm{dm}/f$ $\apprle$ $1.23\times 10^{18}$~yr \citep{boy06}.

We have shown that the reionization history of the universe is sensitive to decaying dark matter parameters, and future experiments may lead to tighter constraints on dark matter models.

\acknowledgments

This work is partially supported by grants from the Research Grant Council of  the Hong Kong Special Administrative Region, China (Project Nos. 400805 and 400910). We thank the ITSC of the Chinese University of Hong Kong for providing its clusters for computations.

\clearpage


\begin{thebibliography}{}
\bibitem[Barger et al.(1995)]{bar95}Barger, V., Phillips, R. J. N. and Sarkar, S. 1995, Phys. Lett. B, 352, 365
\bibitem[Becker et al.(2001)]{bec01}Becker, R. H. et al. 2001, \aj, 122, 2850
\bibitem[Becker et al.(2007)]{bec07}Becker, G. D., Rauch, M., Sargent, W. L. W., 2007, \apj, 662, 72
\bibitem[Biermann \& Kusenko(2006)]{bie06} Biermann, P. L., Kusenko, A., 2006, \prl, 96, 091301
\bibitem[Borzumati et al.(2008)]{bor08}Borzumati, F., Bringmann, T., \& Ullio, P., 2008, \prd, 77, 063514
\bibitem[Boyarsky et al.(2006)]{boy06}Boyarsky, A., Neronov, A., Ruchayskiy, O., \& Shaposhnikov,  M. 2006, \mnras, 370, 213
\bibitem[Cen(2001)]{cen01} Cen, R., 2001, \apj, 546, L77
\bibitem[Chan \& Chu(2011)]{cha10} Chan, M. H., Chu, M. C., 2011, \apj, 727, L47
\bibitem[Dodelson \& Widrow(1994)]{dol94} Dodelson, S., Widrow, L. M., 1994, \prl, 72, 1
\bibitem[Gnedin \& Ostriker(1997)]{gne97}Gnedin, N. Y., Ostriker, J. P. 1997, \apj, 486, 581
\bibitem[Komatsu et al.(2010)]{kom10}Komatsu, E. et al. 2010, \apjs, 192, 18
\bibitem[Lewis et al.(2000)]{lew00}Lewis, A., Challinor, A., \& Lasenby, A. 2000, \apj, 538, 473
\bibitem[Lewis \& Bridle(2002)]{lew02}Lewis, A., Bridle, S. 2002, \prd, 66, 103511
\bibitem[Mapelli \& Ferrara(2005)]{map05} Mapelli, M. \& Ferrara, A. 2005, \mnras, 364, 2
\bibitem[McGreer et al.(2011)]{mcg11} McGreer, I. D., Mesinger, A., Fan, X., 2011, \mnras, 415, 3237
\bibitem[Mitra et al.(2012)]{mit12} Mitra, S., Choudhury, T. R., Ferrara, A., 2012, \mnras, 419, 1480
\bibitem[Osterbrock(1974)]{astrophy-book} Osterbrock, D. E., 1974, Astrophysics of Gaseous Nebulae (W. H. Freeman and Company, San Francisco)
\bibitem[Pandolfi et al.(2011)]{pan11} Pandolfi, S. et al. 2011, arXiv:1111.3570v1
\bibitem[Prokhorov \& Silk(2010)]{pro10} Prokhorov, D. A. and Silk, J. 2010, arXiv:1001.0215
\bibitem[Sasaki \& Umemura(1996)]{sas96}Sasaki, S., \& Umemura, M. 1996, \apj, 462, 104
\bibitem[Seager et al.(2000)]{sea00}Seager, S., Sasselov, D. D., \& Scott, D. 2000, \apjs, 128, 407
\bibitem[Seljak et al.(2006)]{sel06}Seljak, U., Makarov, A., McDonald, P., \& Trac, H. 2006, \prl, 97, 191303
\bibitem[Shull \& van Steenberg(1985)]{shu85}Shull, J. M., van Steenberg, M. E. 1985, \apj, 298, 268
\bibitem[Tegmark et al.(1993)]{teg93}Tegmark, M., Silk, J., \& Evrard, A. 1993, \apj, 417, 54
\bibitem[Zahn et al.(2011)]{zah11}Zahn, O. et al. 2011, arXiv:1111.6386v1
\bibitem[Zhang et al.(2007)]{zha07}Zhang, L., Chen, X., Kamionkowski, M., Si, Z.-G. Si, \& Zheng, Z. 2007, \prd, 76, 061301
\end{thebibliography}
\end{document}